\documentclass[runningheads]{llncs}

\usepackage[T1]{fontenc}
\usepackage[utf8]{inputenc}
\usepackage{amsmath,amssymb,mathtools}
\usepackage{stmaryrd}
\usepackage{booktabs}
\usepackage{enumitem}
\usepackage{graphicx}
\usepackage{listings}
\usepackage{microtype}
\usepackage{needspace}
\usepackage{placeins}
\usepackage{xcolor}
\usepackage{xspace}
\usepackage{hyperref}
\usepackage[nameinlink,noabbrev]{cleveref}

\hypersetup{
  colorlinks=true,
  linkcolor=blue,
  citecolor=blue,
  urlcolor=blue,
  bookmarksopen=true,
  bookmarksnumbered=true,
  bookmarksdepth=2,
  pdftitle={Kairos: Generating Tick-Indexed Proof Obligations for Synchronous Temporal Contracts},
  pdfauthor={Frederic Dabrowski},
  pdfkeywords={deductive verification, temporal contracts, synchronous software, proof obligations, verification tools}
}

\makeatletter
\def\toclevel@paragraph{2}
\makeatother


\setlist[itemize]{topsep=3pt,itemsep=2pt,parsep=0pt,partopsep=0pt}
\setlist[enumerate]{topsep=3pt,itemsep=2pt,parsep=0pt,partopsep=0pt}

\newcommand{\kairos}{\textsc{Kairos}\xspace}
\newcommand{\rocq}{Rocq\xspace}

\newcommand{\spot}{Spot\xspace}

\definecolor{kairospurple}{HTML}{4B3F72}
\definecolor{kairosblue}{HTML}{0B6E8A}
\definecolor{kairosgreen}{HTML}{2F6B3F}
\definecolor{kairosgray}{HTML}{5C6773}
\definecolor{kairosback}{HTML}{F7F8FA}

\lstdefinelanguage{Kairos}{
  morekeywords={
    type,node,returns,observers,spec,def,predicate,init,step,action,
    contracts,requires,ensures,states,invariants,in,except,transitions,to,
    when,if,then,else,end,bool,int,true,false,and,or,not,pre,pre_k,past,G,X,W,
    Formula,HExpr
  },
  morekeywords=[2]{
    Init,Run,Idle,Infusing,Bolus,Alarm,BasalInfusing,BolusDelivering,
    BolusLockoutIdle,BolusLockoutBasal,StopHold,Running,AlarmResponse,
    AlarmLatched
  },
  keywordstyle=[2]\color{kairosblue}\bfseries,
  sensitive=true,
  morecomment=[l]{//},
  morestring=[b]"
}

\lstdefinestyle{kairoscode}{
  language=Kairos,
  basicstyle=\ttfamily\small,
  keywordstyle=\color{kairospurple}\bfseries,
  commentstyle=\color{kairosgray},
  stringstyle=\color{kairosgreen},
  identifierstyle=\color{black},
  numbers=left,
  numberstyle=\scriptsize\color{kairosgray},
  numbersep=6pt,
  backgroundcolor=\color{kairosback},
  frame=single,
  framesep=4pt,
  rulecolor=\color{black!20},
  columns=fullflexible,
  keepspaces=true,
  showstringspaces=false,
  breaklines=true,
  lineskip=0.25pt,
  aboveskip=0.5\baselineskip,
  belowskip=0.5\baselineskip,
  tabsize=2
}

\lstdefinestyle{kairosfullcode}{
  style=kairoscode,
  basicstyle=\ttfamily\fontsize{7.4pt}{7.1pt}\selectfont,
  numbers=none,
  aboveskip=0.25\baselineskip,
  belowskip=0.25\baselineskip
}

\title{Kairos: Generating Tick-Indexed Proof Obligations for Synchronous Temporal Contracts}
\titlerunning{Kairos Tick-Indexed Proof Obligations}

\author{Fr\'ed\'eric Dabrowski}
\authorrunning{F. Dabrowski}
\institute{Univ. Orl\'eans, INSA Centre Val de Loire, LIFO UR 4022, Orl\'eans, France\\
\email{frederic.dabrowski@univ-orleans.fr}}

\begin{document}
\maketitle

\begin{abstract}
Requirements for synchronous programs relate observations across ticks, while
general-purpose deductive backends use first-order conditions on local steps.
Kairos is a prototype toolchain generating such obligations from synchronous
programs and source-level temporal safety contracts in assume-guarantee form.
Using the standard automata-theoretic reduction, it translates their linear
temporal logic (LTL) formulas into bad-state automata, forms their product with
the program, and derives tick-indexed Why3 obligations. We formalize the
underlying principles and prove reactive contract correctness: under the stated program and automaton
hypotheses, every input trace satisfying the environment assumptions induces a
unique execution satisfying the temporal guarantees.
The definitions, obligation construction, and proof are mechanized in Rocq. A
medical infusion controller illustrates the workflow.
\end{abstract}

\section{Introduction}
\label{sec:introduction}

Synchronous reactive programs execute in logical ticks: each tick samples
inputs, updates internal state, and publishes outputs. Their requirements are
naturally temporal: an alarm should remain latched until it is acknowledged, a
motor should stay off while a hazard is active, and displayed values should
track accumulated histories. Classical automata-theoretic verification
represents temporal specifications by automata and reasons over a product with
the program transition system \cite{vardi-wolper-1986,alpern-schneider-1989}.
For safety properties, the proof can be organized as an invariant-style
exclusion of bad product states \cite{alpern-schneider}.

Connecting this invariant argument to a general-purpose deductive backend
requires transforming product transitions into first-order pre/postconditions
for program commands. This transformation raises several related challenges:
one product source may admit several compatible guarantee successors,
proved safe incoming cases must be propagated as a disjunction to the
obligations generated from their destination,
source and destination annotations describe histories before and after the
command, so their historical reads require explicit shifts, and assumption-bad
and guarantee-bad successors impose different proof requirements. The result
must be a finite family of first-order obligations that covers every compatible
automaton successor and preserves the semantics of historical assertions.

\kairos realizes this bridge for a synchronous source language. Users write
source-level assume/guarantee safety contracts over current and historical
values. The prototype translates them to safety automata and synchronizes the
automata with normalized program ticks. It then groups the resulting product
cases by normalized tick, product source, and assumption successor, retaining
every compatible guarantee successor in guarded disjunctive postconditions,
transporting historical assertions between tick endpoints, and emitting
preservation or bad-guarantee exclusion tasks. Why3 then receives first-order
obligations around the actual command body \cite{why3}.

Under the stated program and automaton hypotheses, validity of the generated
obligations entails reactive correctness: every input trace satisfying the
assumptions induces a unique execution satisfying the temporal guarantees. The
argument, mechanized in Rocq, connects these local obligations to the trace-level
result.

\smallskip
\noindent\textit{Contributions.} We contribute the \kairos language and
end-to-end toolchain, illustrated by a medical infusion controller; the
tick-indexed obligation construction; and its mechanized reactive-correctness
theorem. The
\href{https://github.com/DabrowskiFr/Kairos}{implementation} and
\href{https://github.com/DabrowskiFr/vsste26-rocq-artefact}{Rocq formalization}
are available online.

\section{Running Example}
\label{sec:running-example}

This section introduces the \kairos language through a medical infusion
controller represented as a \emph{node}, the unit compiled and verified by the
tool. At each tick, it samples an infusion request
(\texttt{infusionRequested}), acknowledgment of a latched alarm (\texttt{ack}),
occlusion signal (\texttt{occlusion}), requested rate (\texttt{rate}), volume
delivered during the preceding sampling interval (\texttt{delivered}), and
prescription- or device-configured limits (\texttt{maxRate},
\texttt{doseLimit}). It outputs the next-interval motor command
(\texttt{motorOn}), latched alarm state (\texttt{alarmLatched}), and cumulative
delivered dose (\texttt{totalDose}). \Cref{lst:medical-example} shows a compact
fragment centered on the alarm-response transition and its immediate-response
and weak-until guarantees, followed below. It omits predicates and branches supporting admission,
initialization, flow checks, and alarm release, whose relevant roles are
described below.

Node contracts are safety LTL formulas over tick-observation sequences, with
atomic propositions given by first-order assertions whose historical expressions
may read current values or bounded history. Besides standard Boolean
connectives, \texttt{X}~$\varphi$ requires $\varphi$ at the next tick,
\texttt{G}~$\varphi$ at every tick, and $\varphi~\texttt{W}~\psi$ requires
$\varphi$ until $\psi$, or forever if $\psi$ never occurs. Within atoms,
\texttt{pre(x)} reads \texttt{x} one tick back and \texttt{pre\_k(x,k)}
$k$ ticks back; every historical read must be initialized on each path
to its use. The listing uses one-step history in the observer recurrence and a
temporal predicate; deeper reads are supported but unnecessary here.

\noindent\begin{minipage}{\linewidth}
\begin{lstlisting}[style=kairoscode,caption={Medical infusion controller fragment used in the running example.},label={lst:medical-example}]
spec def holds_until_release(trigger: Formula, hold: Formula,
  clear: Formula) = X G ($trigger => X($hold W $clear));
node medical_infusion_light(
  infusionRequested: bool, occlusion: bool, ack: bool,
  rate: int, delivered: int, maxRate: int, doseLimit: int)
returns (motorOn: bool, alarmLatched: bool, totalDose: int)
observers
  measuredDose : int = delivered -> pre(measuredDose) + delivered;
predicate hazardActive() = occlusion = true;
predicate alarmCondition() = hazardActive() or rate > maxRate
  or totalDose + delivered > doseLimit or unexpectedFlow();
predicate doseWouldExceedAtTickEntry() =
  pre(totalDose) + delivered > doseLimit;
predicate alarmConditionAtTickEntry() =
  hazardActive() or rate > maxRate or doseWouldExceedAtTickEntry()
  or unexpectedFlowAtTickEntry();
predicate clearAlarmAtTickEntry() =
  ack = true and not alarmConditionAtTickEntry();
// omitted: admission, initialization, flow, and executable release
action finishTick() { totalDose := totalDose + delivered; }
contracts
  requires: G (rate >= 0 and delivered >= 0
    and maxRate >= 0 and doseLimit >= 0);
  ensures: G ((hazardActive() or rate > maxRate) =>
    alarmLatched = true and motorOn = false);
  ensures: holds_until_release([alarmLatched = true and ack = false],
    [alarmLatched = true and motorOn = false],[clearAlarmAtTickEntry()]);
// omitted: dose, display, flow, latch, and bounded-operation guarantees
states
  Init(init), Idle, Running, AlarmResponse, AlarmLatched;
invariants
  in states except {Init}: totalDose = measuredDose;
  in states except {Init, Running}: motorOn = false;
  in {AlarmResponse, AlarmLatched}: alarmLatched = true;
transitions
  // omitted: Init, Idle, and AlarmResponse branches
  Running:
    to AlarmResponse when alarmCondition()
      { motorOn, alarmLatched := false, true; finishTick(); }
    to Running when admittedNow() { finishTick(); }
    to Idle { motorOn, alarmLatched := false, false; finishTick(); }
  AlarmLatched:
    to AlarmResponse when alarmCondition() { finishTick(); }
    to Idle when clearAlarm() { motorOn, alarmLatched := false, false; }
end
\end{lstlisting}
\end{minipage}

Lines~1--2 define \texttt{holds\_until\_release}. Its \texttt{Formula}
parameters are instantiated by bracketed call-site formulas, whose occurrences
are marked by \texttt{\$}. Specification definitions may also accept historical
expressions or expansion-time natural numbers, for example to set
\texttt{pre\_k} depths.

Lines~3--6 declare the node interface, separating sampled environment inputs
from controller-maintained outputs; nodes may also declare persistent locals,
absent here.

Lines~7--20 contain constructs elaborated before obligation generation. The
specification-only observer defines cumulative delivery: \texttt{measuredDose}
is initially \texttt{delivered} and thereafter its previous value plus the
current sample. Predicates are pure Boolean macros: executable uses are
evaluated without history at transition entry, whereas contract uses are
evaluated on tick observations and may contain \texttt{pre}. Lines~12--16
reconstruct entry dose overflow and combine it with
occlusion, excessive rate, and positive delivery after
\texttt{pre(motorOn)} was false; lines~17--18 require \texttt{ack} and negate
that condition for release. Omitted predicates handle admission (a request, no
alarm or hazard, in-range values, and either a previously active motor or zero
delivery), first-tick initialization, and the corresponding executable release.
\texttt{finishTick} expands to \texttt{totalDose := totalDose + delivered} in
each transition body. Elaboration expands definitions and macros, lowers
observers to proof-only variables updated after transitions, and yields
contracts over a finite control graph.

Lines~21--28 declare contracts: \texttt{requires} states an input-only
environment assumption, under which each \texttt{ensures} is proved. The shown
guarantees demand a latched alarm and stopped motor on occlusion or excessive
rate, and maintain that response after an unacknowledged alarm until an
acknowledged alarm-free release (or forever). Omitted guarantees require
displayed and observed cumulative dose to agree from the next tick onward, and
cover unexpected delivery, alarm latching, and dose bounds.

Lines~29--34 declare finite control, with \texttt{Init(init)} initial, and attach
first-order invariants to states selected by \texttt{in}. These are local proof
annotations, not global assumptions: obligations may use the source annotation
and must establish the destination annotation. Since destination invariants are
shifted into command postconditions, they may read past but not current input
samples. The three shown annotations relate display to observer, keep the motor
off outside infusion, and characterize alarm states.

Lines~35--45 show selected transitions. Omitted \texttt{Init} branches test
initial alarm and admission, initialize outputs, and enter the main control
states; \texttt{Idle} likewise prioritizes alarm over admission and otherwise
self-loops with the motor off. \texttt{AlarmResponse} branches handle a fresh
alarm or enter \texttt{AlarmLatched}.

In \texttt{to S when g \{ c \}}, selecting \(g\) moves control to \(S\) and
executes \(c\). Command bodies support assignments, \texttt{skip}, conditionals,
matches, and annotated \texttt{while} loops. Source order sets priority; if no
branch is enabled at a state, including one with no declared transitions,
normalization adds an empty self-step. An unguarded final branch is therefore
used as the default, as in \texttt{Running}, and multiple assignment updates both
outputs together.

Operationally, each logical tick starts from stored control, outputs, and locals,
then samples fresh inputs. The selected transition atomically updates control
and store; the visible observation combines sampled inputs with resulting
outputs, while unassigned outputs and locals persist. Iterating this reaction
over successive samples of an input stream yields the node's visible input/output
trace.

\FloatBarrier

\section{Obligation-Generation Pipeline}
\label{sec:tool-pipeline}

This section follows the medical controller's
\texttt{Running}-to-\texttt{AlarmResponse} transition through the executable
pipeline, from source-node normalization and product summarization to the finite
first-order contracts submitted to Why3, and provides intuition for the
construction proved correct in \Cref{sec:proof-interface}.

\paragraph{From source ticks to product summaries.}
Obligation construction starts from normalized transitions with a priority-aware
guard and atomic body.
First, the source assumptions are conjoined into a single requirement \(R\).
\spot checks that \(R\) and each \(G_i\) denote safety properties, then builds
one complete assumption automaton \(A_R\) and one complete guarantee automaton
\(A_{G_i}\) for each \(G_i\) \cite{spot}. Completeness means that every state has an
outgoing edge for every tick observation. For each \(i\), \kairos forms the
synchronized product \(\mathcal P\times A_R\times A_{G_i}\). A bad state records
a violating prefix.
Assumption-bad successors impose no guarantee obligation and are discarded.
From active product sources, whose assumption and guarantee components are both
non-bad, \kairos groups the remaining edges by product source, normalized
program transition, and assumption edge, producing one summary per group. A
summary may therefore contain several guarantee edges for
the same program transition. Intuitively, the guarantee automaton may admit
several continuations for that reaction; their guards identify those compatible
with the resulting observation, and verification must establish that at least one
is enabled.

\Needspace{8\baselineskip}
For illustration, let \(C\) denote the predicate
\texttt{clearAlarmAtTickEntry}, and let
\(T\equiv\texttt{alarmLatched}\land\neg\texttt{ack}\),
\(D\equiv\texttt{occlusion}\lor\texttt{rate}>\texttt{maxRate}\), and
\(H\equiv\texttt{alarmLatched}\land\neg\texttt{motorOn}\).
We consider only the guarantees
\(G_W\equiv\mathtt{X}\mathtt{G}(T\Rightarrow
\mathtt{X}(H\mathbin{\mathtt{W}}C))\) and
\(G_A\equiv\mathtt{G}(D\Rightarrow H)\). Space precludes displaying
\(\mathcal P\times A_R\times A_{G_W}\) and
\(\mathcal P\times A_R\times A_{G_A}\) in full. The automata
\(A_R,A_{G_W},A_{G_A}\) respectively have states
\(A_0,A_{\mathsf{bad}}\),
\(G^W_0,G^W_1,G^W_2,G^W_{\mathsf{bad}}\), and
\(G^A_0,G^A_{\mathsf{bad}}\); the \(0\)-states are initial, the explicitly
marked states are bad, and all others non-bad.
For the program transition
\(\tau_A:\texttt{Running}\rightarrow\texttt{AlarmResponse}\), grouping the
induced edges yields summaries \(\sigma_W\) and \(\sigma_A\). Both use
\(e_R:A_0\xrightarrow{\alpha}A_0\). Their respective source keys are
\(\kappa_W=(\texttt{Running},A_0,G^W_2)\) and
\(\kappa_A=(\texttt{Running},A_0,G^A_0)\). Their guarantee-edge projections,
partitioned into safe and \(\mathsf{bad}_G\) successors, are
\[
\begin{array}{ll}
\sigma_W: & \mathsf{safe}:\ 
G^W_2\xrightarrow{\gamma_a}G^W_2,\quad
G^W_2\xrightarrow{\gamma_p}G^W_1;\\
\sigma_A: & \mathsf{safe}:\ G^A_0\xrightarrow{\gamma_s}G^A_0,\qquad
\mathsf{bad}_G:\ G^A_0\xrightarrow{\gamma_b}G^A_{\mathsf{bad}}.
\end{array}
\]
For readability, we show source-level formulas equivalent to the emitted expanded
guards. The
assumption edge has guard
\(\alpha\equiv\texttt{rate}\geq 0\land\texttt{delivered}\geq 0
\land\texttt{maxRate}\geq 0\land\texttt{doseLimit}\geq 0\). The weak-until edges use
\(\gamma_a\equiv\texttt{ack}\lor\neg\texttt{alarmLatched}\) and
\(\gamma_p\equiv\neg\texttt{ack}\land\texttt{alarmLatched}\); the \(G_A\)
edges use \(\gamma_s\equiv\neg D\lor H\) for satisfaction and
\(\gamma_b\equiv D\land\neg H\) for violation.

\paragraph{From the summaries to proof obligations.}
For \(\sigma\in\{\sigma_W,\sigma_A\}\), the generated precondition has the
same abstract shape
\[
\mathsf{Pre}_\sigma =
U_{\texttt{Running}}\land\mathsf{Incoming}(\kappa_\sigma)
\land B_\sigma\land\alpha\land\mathsf{EntryCoherence}.
\]
\(U_{\texttt{Running}}\) is the current user annotation, including
\(\texttt{totalDose}=\texttt{measuredDose}\).
\Cref{subsec:obligation-construction} defines \(\mathsf{Incoming}\) exactly.
For both \(\kappa_W\) and \(\kappa_A\), it is a disjunction of safe
\texttt{Init}, \texttt{Idle}, and \texttt{Running} predecessor cases. At source
level, each case says that the preceding tick admitted infusion and reached
\texttt{Running} with
\(\neg\texttt{alarmLatched}\land\texttt{totalDose}=\texttt{measuredDose}\).
The cases differ in predecessor control state and in the transported source
annotation, program and guarantee guards, sampled-input assumption, and body
equations. The remaining factors concern
the current tick:
\(B_\sigma\) represents the normalized \texttt{alarmCondition()} guard at tick
entry, \(\alpha\) is the assumption-edge guard defined above, and
\(\mathsf{EntryCoherence}\) enforces \(y=\texttt{pre}(y)\) for every stored
non-input \(y\).

The outgoing guarantee cases determine the postconditions:
\(\mathsf{SomeSafe}_{\sigma_W}=\gamma_a\lor\gamma_p\),
\(\mathsf{SomeSafe}_{\sigma_A}=\gamma_s\), and
\(\mathsf{NoBad}_{\sigma_A}=\neg\gamma_b\). For \(X\in\{W,A\}\), the resulting
postconditions have the abstract shapes
\[
\mathsf{Post}^+_{\sigma_X}=\mathsf{SomeSafe}_{\sigma_X}
\land\mathsf{Inv}^{\mathsf{post}}_{\texttt{AlarmResponse}}
\land\mathsf{Contrib}^{\mathsf{post}}_{\sigma_X},\qquad
\mathsf{Post}^-_{\sigma_A}=\mathsf{NoBad}_{\sigma_A}.
\]

Safe successors share the \texttt{AlarmResponse} annotation
\(H\land\texttt{totalDose}=\texttt{measuredDose}\), yielding
\(\mathsf{Inv}^{\mathsf{post}}_{\texttt{AlarmResponse}}\). Its transport is
trivial here; generally, past reads move one boundary toward the present and
current-input reads are excluded because the destination sample is unavailable.
For \(\sigma\in\{\sigma_W,\sigma_A\}\), each safe edge \(e\) targeting a
noninitial key contributes to \(\mathsf{Contrib}^{\mathsf{post}}_\sigma\) a
\(\gamma_e\)-guarded record transporting \(U_{\texttt{Running}}\) and
\(B_\sigma\) to the post-state and retaining \(\alpha\), \(\gamma_e\), and
\(\beta_\sigma\), the body-effect formula. The soundness result is parametric in
\(\beta_\sigma\): any
strengthening proved by the safe obligation may be propagated, and
\(\top\) is always admissible. The prototype proposes \(\beta_\sigma\) by
straight-line symbolic execution, adding frame equalities and forgetting
variables modified by unsupported compound statements. Thus \(\sigma_W\) records
\(\gamma_a\) and \(\gamma_p\) for
\(G^W_2\) and \(G^W_1\), while \(\sigma_A\) records \(\gamma_s\) for \(G^A_0\);
the destination annotation is the separate
\(\mathsf{Inv}^{\mathsf{post}}\) conjunct. The common body \(c_{\tau_A}\)
establishes
\(H\), increments both \texttt{totalDose} and \texttt{measuredDose} by
\texttt{delivered}.
For \(\sigma_W\), \(\gamma_a\lor\gamma_p\) reduces to
\(\texttt{ack}\lor\neg\texttt{ack}\), establishing safe progress; for
\(\sigma_A\), \(H\) establishes \(\gamma_s\) and excludes \(\gamma_b\).
Together, these facts establish \(\mathsf{SomeSafe}\), which asserts the
existence of a compatible non-bad guarantee successor;
\(\mathsf{Contrib}^{\mathsf{post}}\) instead carries proved path facts into later
preconditions.

The resulting formulas yield the positive contracts for both summaries and
the alarm-response exclusion contract around the same body:
\[
\{\mathsf{Pre}_{\sigma_W}\}\ c_{\tau_A}\ \{\mathsf{Post}^+_{\sigma_W}\},
\qquad
\{\mathsf{Pre}_{\sigma_A}\}\ c_{\tau_A}\ \{\mathsf{Post}^+_{\sigma_A}\},
\qquad
\{\mathsf{Pre}_{\sigma_A}\}\ c_{\tau_A}\ \{\mathsf{Post}^-_{\sigma_A}\}.
\]
The semantic safety goal excludes any product step to a guarantee-bad state
while the assumption remains non-bad. Inductively, each safe obligation proves
a compatible safe successor and a sound post-state record; transporting that
record into the destination characteristic strengthens later local
preconditions. Thus every propagated fact derives from a proved safe-predecessor
obligation.

\paragraph{From proof obligations to Why3.}
The backend lowers the remaining symbolic \texttt{pre}/\texttt{pre\_k} reads to
immutable WhyML parameters for source-history slots. The obligations quantify
universally over these parameters, which represent the considered tick history.
Their discharge proves total
correctness of the normalized body, justifying the total statement denotation
used in \Cref{sec:proof-interface}.

\paragraph{Generated artifacts.}
The following table reports artifacts generated for the accepted medical
programs.
The light case is the running example; the full case adds patient-controlled boluses with
remaining-dose memory, post-bolus lockout, explicit stop handling, another setup
hazard, and richer display/audit histories for peak rate and classified
delivered dose. Proof timings use 10 parallel jobs and a 20\,s per-goal timeout.
LoC counts non-comment lines; St., Tr., A, and G count states, transitions,
assumptions, and guarantees. Prod./Summ. count product states and canonical
intermediate-representation summaries across all guarantees; summaries precede backend
grouping, whereas Safe/BadG count the resulting Why3 goals after grouping and
splitting. Valid gives discharged/total goals; Wall gives elapsed \kairos time
in seconds.

\begin{center}
\scriptsize
\begin{tabular}{@{}lrrrrrrrrrr@{}}
\toprule
Case & LoC & St. & Tr. & A/G & Prod. & Summ. & Safe & BadG & Valid & Wall(s) \\
\midrule
medical light & 81 & 5 & 12 & 1/7 & 34 & 95 & 17 & 69 & 86/86 & 0.9 \\
medical full & 502 & 9 & 43 & 2/21 & 186 & 882 & 56 & 638 & 694/694 & 32.3 \\
\bottomrule
\end{tabular}
\end{center}

Both medical examples are available in the repository, together with further
examples and counterexamples used for regression testing.

\section{Correctness of Generated Obligations}
\label{sec:proof-interface}

This section establishes the correctness of the mathematical construction
underlying the obligations of \Cref{sec:tool-pipeline}. The proof
has two layers.  A standard bad-prefix argument reduces contract validity to a
local invariant certificate.  The \kairos-specific layer instantiates this
argument with the normalized statement core, constructs its summaries and
deductive obligations, and proves that their validity provides such a
certificate.
Each layer states its own assumptions: automata must expose violations,
normalized programs must react deterministically without changing inputs, and
generated formulas must encode product cases at available history depths.

\subsection{Product semantics and invariant certificate}

At the semantic level, we represent a program by a labelled transition system
\(\mathcal P=(S,s_0,O,\xrightarrow{\cdot})\), where \(S\) is its state space,
\(s_0\in S\) its initial state, \(O\) its observation space, and
\(\xrightarrow{\cdot}\subseteq S\times O\times S\) its labelled transition
relation. Intuitively, \(s\xrightarrow{o}s'\) represents one logical tick from
\(s\) to \(s'\), and \(o\) carries the semantic data on which contract guards
are evaluated; the normalized instance below exposes inputs and outputs through
a projection of this label.
An observation automaton over \(U\) is
\(A=(Q,q_0,\mathsf{bad},\delta)\), where \(Q\) is its state space,
\(q_0\in Q\) its initial state, \(\mathsf{bad}\subseteq Q\) its set of bad
states, and \(\delta\subseteq U\times Q\times Q\) its transition relation.
It is complete when \(\forall u,q.\;\exists q'.\;\delta(u,q,q')\).
For \(w\in U^\omega\), let \(\mathsf{Covered}_A(w)\) mean that every finite
prefix of \(w\) labels a run from \(q_0\), and let
\(\mathsf{BadReach}_A(w)\) mean that some such run reaches \(\mathsf{bad}\).
The induced safety semantics on infinite words is
\[
 w\Vdash A\quad\Longleftrightarrow\quad
 \mathsf{Covered}_A(w)\land\neg\mathsf{BadReach}_A(w).
\]
The coverage conjunct is essential: absence of a reachable bad state alone
would let a partial automaton accept by blocking.
Both contract automata read \(O\). Since safety is closed under conjunction, the
formalization uses one \(A_G\) for all guarantees; the implementation processes
them separately to avoid a larger conjunctive automaton and product.

Synchronizing the program and automata makes a contract violation a reachable
product state with a non-bad assumption and bad guarantee component. For
supplied automata
\(A_R=(Q_R,r_0,\mathsf{bad}_R,\delta_R)\) and
\(A_G=(Q_G,g_0,\mathsf{bad}_G,\delta_G)\), product states are
\(x=(s,r,g)\), and the initial state is \(x_0=(s_0,r_0,g_0)\).
\paragraph{Standing automaton assumptions.}
Assume non-bad initial states, complete \(A_G\), and absorbing \(A_R\)-bad
states:
\(\mathsf{bad}_R(r)\land\delta_R(o,r,r')\Rightarrow
\mathsf{bad}_R(r')\). These support initialization, trace coverage, and
invariant soundness; finite edge lists and trace characterization appear below.
The product synchronizes them on the observation labeling the program step:
\[
  (s,r,g)\xRightarrow{o}(s',r',g')
  \quad\Longleftrightarrow\quad
  s\xrightarrow{o}s'
  \land\delta_R(o,r,r')
  \land\delta_G(o,g,g').
\]
We distinguish active states from guarantee-bad states:
\[
  \mathsf{Active}(s,r,g)\triangleq
    \neg\mathsf{bad}_R(r)\land\neg\mathsf{bad}_G(g),
  \qquad
  \mathsf{BadG}(s,r,g)\triangleq
    \neg\mathsf{bad}_R(r)\land\mathsf{bad}_G(g).
\]
Writing \(\xRightarrow{*}\) for product reachability, with edge labels
existentially hidden, define
\[
\mathsf{ContractValid}(\mathcal P,A_R,A_G)
\triangleq
\forall x.\;x_0\xRightarrow{*}x\Rightarrow\neg\mathsf{BadG}(x).
\]
Along a product path, absorption means that a non-bad assumption state rules out
an earlier bad assumption state on that path. No analogous condition is required
for \(A_G\), since reaching a guarantee-bad state while the assumption is
non-bad already yields a violating prefix.

To establish \(\mathsf{ContractValid}(\mathcal P,A_R,A_G)\) inductively, we
use a predicate \(I\) over product states to record the state facts carried
between active states. It is a certificate when, for every edge label \(o\),
it satisfies
\[
\begin{array}{ll}
\textsc{Init}:&
 I(x_0)\land\neg\mathsf{BadG}(x_0),\vphantom{x\xRightarrow{o}y}\\
\textsc{Pres}:&
 I(x)\land\mathsf{Active}(x)\land x\xRightarrow{o}y
 \land\mathsf{Active}(y)
 \Rightarrow I(y),\\
\textsc{Excl}:&
 I(x)\land\mathsf{Active}(x)\land x\xRightarrow{o}y
 \land\mathsf{BadG}(y)
 \Rightarrow\bot.
\end{array}
\tag{1}\label{eq:invariant-certificate}
\]
The non-bad \textsc{Init} conjunct follows from the standing assumptions and is
not generated. Thus \textsc{Pres} handles active successors, \textsc{Excl}
rules out guarantee-bad successors, and assumption-bad successors require no
condition.

\begin{theorem}[Invariant soundness]
\label{thm:invariant-soundness}
Every predicate satisfying \eqref{eq:invariant-certificate} establishes
\(\mathsf{ContractValid}(\mathcal P,A_R,A_G)\).
\end{theorem}

Alpern and Schneider combine invariants for safety with candidate and variant
functions for liveness over Boolean combinations of B\"uchi
automata \cite{alpern-schneider-1989}. Our theorem retains only the
safety-invariant component of their method and specializes the invariant domain
to the combined state of a program with separate assumption and guarantee
safety automata. This specialization yields an asymmetric proof rule:
assumption-bad successors carry no obligation, whereas guarantee-bad successors
are excluded while the assumption remains non-bad. The following subsections
connect this principle to the toolchain by defining the normalized tick
semantics and proving that the generated first-order obligations establish the
required certificate.

\subsection{Normalized programs and tick semantics}

We now instantiate \(\mathcal P\) with normalized tick semantics, exposing the
endpoints used by generated assertions. Fix sets \(\mathsf{Var}\) and \(V\) of
variables and values. A store is a
valuation \(\rho\in\mathsf{Store}=\mathsf{Var}\to V\).
An executable expression is modeled extensionally as an element of
\(\mathsf{Expr}=\mathsf{Store}\to V\), and a guard as an element of
\(\mathsf{Guard}=2^{\mathsf{Store}}\). A statement is represented
extensionally by a total state transformer
\(\mathsf{Stmt}=\mathsf{Store}\to\mathsf{Store}\). We use
\(e\in\mathsf{Expr}\), \(b\in\mathsf{Guard}\), and \(c\in\mathsf{Stmt}\) as
metavariables; \(e(\rho)\) and \(c(\rho)\) denote expression evaluation and
statement execution.
Modeling executable expressions and statements as total functions simplifies
the temporal proof. \kairos accepts a program only after Why3 has established,
under each generated transition precondition, that the corresponding body is
defined, fault-free, and terminating; totality is therefore not an additional
assumption for accepted programs.

To interpret transported assertions, memories contain both a persistent store and
its history. Let
\(\mathsf{Hist}=(\mathsf{Var}\times\mathbb N)\to V\), and let \(M\) be the set of memories
\(m=(\rho_m,\mathsf{past}_m)\), with
\(\rho_m\in\mathsf{Store}\) and
\(\mathsf{past}_m\in\mathsf{Hist}\).
For \(J\subseteq\mathsf{Var}\), let
\(\mathsf{Coherent}_J(m)\triangleq
 \forall v\notin J.\ \rho_m(v)=\mathsf{past}_m(v,0)\).
A normalized program is a tuple
\(\mathcal C=(\mathsf{In},\mathsf{Out},\allowbreak Q,q_0,m_0,T)\), where
\(\mathsf{In},\mathsf{Out}\subseteq\mathsf{Var}\) are its named input and
output variables, \(Q\) is a set of control states, \(q_0\in Q\) is the initial
control state, \(m_0\in M\) satisfies
\(\mathsf{Coherent}_{\mathsf{In}}(m_0)\), and
\(T\subseteq Q\times\mathsf{Guard}\times\mathsf{Stmt}\times Q\)
is a finite set of transitions.
The condition on \(m_0\) only initializes this representation invariant.
We write \(\tau=(q,b,c,q')\in T\) for a transition. Its configurations
are pairs \(s=(q,m)\in S=Q\times M\), with
\(s_{\mathcal C}^0=(q_0,m_0)\) initially.
An input sample is a function \(\iota:\mathsf{In}\to V\), and an output sample
is a function \(u:\mathsf{Out}\to V\). A memory \(m\) and an input sample
\(\iota\) determine
the valuation \(\rho_{m,\iota}\), obtained by installing \(\iota\) in
\(\rho_m\): input variables take their sampled values and other variables
retain their current values. For source memory \(m\) and destination memory
\(m'\), define the entry and post histories by
\(H_{\mathsf e}^{m,\iota}(x,0)=\rho_{m,\iota}(x)\),
\(H_{\mathsf p}^{m,m'}(x,0)=\rho_{m'}(x)\), and, for \(k\geq0\),
\(H_{\mathsf e}^{m,\iota}(x,k+1)=H_{\mathsf p}^{m,m'}(x,k+1)
=\mathsf{past}_m(x,k)\). Let
\(O=(\mathsf{In}\to V)\times(\mathsf{Out}\to V)\times
\mathsf{Hist}\times\mathsf{Hist}\). A semantic tick label is
\(o=(\iota,u,H_{\mathsf e}^{m,\iota},H_{\mathsf p}^{m,m'})\); write
\(H_{\mathsf e}(o)\) and \(H_{\mathsf p}(o)\) for its history components and
\(\mathsf{vis}(o)=(\iota,u)\) for its visible I/O projection. These histories
are logical views of the memories, not additional runtime state. The tick rule
uses two boundary predicates:
\(\mathsf{InputStable}\) preserves the sampled inputs, while
\(\mathsf{Complete}\) installs the resulting store and advances its history.
For \(\tau=(q,b,c,q')\in T\) and the label \(o\) above, define
\[
\begin{gathered}
 \begin{aligned}
  (q,m)\xrightarrow[\tau]{o}(q',m')
  \quad\Longleftrightarrow\quad&
  \rho_{m,\iota}\in b\ \land\
  \mathsf{InputStable}(\rho_{m,\iota},c(\rho_{m,\iota}))\\
  &{}\land\ \mathsf{Complete}(m,c(\rho_{m,\iota}),m')
  \ \land\ u=\rho_{m'}|_{\mathsf{Out}}
 \end{aligned}\\[0.5ex]
 \begin{aligned}
  \text{where}\quad \mathsf{InputStable}(\rho,\rho')
  &\triangleq \forall x\in\mathsf{In}.\ \rho'(x)=\rho(x),\\
  \phantom{\text{where}\quad}\mathsf{Complete}(m,\rho',m')
  &\triangleq \rho_{m'}=\rho'
    \land \forall x.\ \mathsf{past}_{m'}(x,0)=\rho'(x)\\
  &\quad{}\land \forall x,k.\ \mathsf{past}_{m'}(x,k+1)=\mathsf{past}_m(x,k).
 \end{aligned}
\end{gathered}
\]
The predicate \(\mathsf{Complete}\) establishes
\(\mathsf{Coherent}_{\mathsf{In}}(m')\); consequently this property holds at
every reachable tick boundary.
This rule induces the transition system
\(\mathcal P=\llbracket\mathcal C\rrbracket_{\mathsf{tick}}\):
\(s\xrightarrow{o}s'\) holds exactly when the rule above holds for some
\(\tau\in T\).
A run is
\(\varrho=((s_n)_{n\in\mathbb N},(o_n)_{n\in\mathbb N})\) with
\(s_0=s_{\mathcal C}^0\) and \(s_n\xrightarrow{o_n}s_{n+1}\) for every \(n\).
Let \(\mathsf{Runs}(\mathcal P)\) be the set of runs and
\(\mathsf{tr}(\varrho)=o_0o_1\ldots\) its semantic tick-label trace. For an input
stream \(\bar\iota\in(\mathsf{In}\to V)^\omega\),
\(\mathsf{Runs}(\mathcal P,\bar\iota)\) contains the runs whose labels
have input projection \(\bar\iota\); applying \(\mathsf{vis}\)
pointwise yields their visible input/output traces.

Run existence and uniqueness additionally require three structural conditions.
Let
\(\mathsf{Cover}\) abbreviate
\(\forall(q,m),\iota.\ \exists(q,b,c,q')\in T.\
\rho_{m,\iota}\in b\), while
\(\mathsf{InputReadOnly}\) abbreviates
\(\forall(q,b,c,q')\in T,\rho,x\in\mathsf{In}.\
c(\rho)(x)=\rho(x)\). Finally, \(\mathsf{Det}\) requires, for all
\((q,m),\iota\) and \((q,b_j,c_j,q'_j)\in T\), \(j\in\{1,2\}\), that
\(\rho_{m,\iota}\in b_1\cap b_2\) imply
\((q'_1,c_1(\rho_{m,\iota}))=(q'_2,c_2(\rho_{m,\iota}))\).
\paragraph{Standing normalized-core assumptions.}
The model assumes
\(\mathsf{Cover}\), \(\mathsf{InputReadOnly}\), and \(\mathsf{Det}\), ensured in
\kairos cores by complementary self-loops, frontend input-read-only validation,
and priority-disjoint guards, respectively.

\subsection{Construction of deductive obligations}
\label{subsec:obligation-construction}

To derive the certificate \eqref{eq:invariant-certificate}, we align historical
endpoints, summarize every active product edge, and turn safe and bad summaries
into preservation and exclusion triples.

We first fix the syntax and admissible depths of generated historical assertions.
Fix a normalized program \(\mathcal C\) and let
\(\mathcal P=\llbracket\mathcal C\rrbracket_{\mathsf{tick}}\). Each memory \(m\)
represents the total history \(H_m\in\mathsf{Hist}\), with
\(H_m(x,0)=\rho_m(x)\) and
\(H_m(x,k+1)=\mathsf{past}_m(x,k)\); the frontier below excludes uninitialized
reads.
Generated historical assertions range over \(\Phi\), defined by
\[
 t ::= v@k,\qquad
 \varphi ::= \top\mid\bot\mid A(\bar t)\mid\neg\varphi
 \mid\varphi\land\varphi\mid\varphi\lor\varphi
 \mid\varphi\Rightarrow\varphi .
\]
Here \(v\in\mathsf{Var}\), \(k\in\mathbb N\), \(\bar t\) is a finite term
sequence, and \(A\) is an arbitrary pure predicate over its values. For
\(H\in\mathsf{Hist}\), \(\llbracket v@k\rrbracket_H=H(v,k)\); satisfaction
\(H\models\varphi\) is standard for atoms and Boolean connectives. Thus
\(v@0\) denotes the current value, \(v@1\) corresponds to
\texttt{pre(v)}, and positive depths represent the source \texttt{pre\_k}
reads. Let \(d(\varphi)\) be zero for read-free \(\varphi\), otherwise its
maximum read depth. An initialization frontier assigns lower-bound ages
\(a(\kappa)\) to product keys \(\kappa=(q,r,g)\);
\(\kappa_0=(q_0,r_0,g_0)\) has age \(0\), and each key edge satisfies
\(a(\kappa')\leq a(\kappa)+1\);
hence \(a(\kappa)\leq n\) on any \(n\)-edge path to \(\kappa\). Every control
annotation, normalized guard, automaton edge label, and predecessor record
\(\varphi\) used at \(\kappa\) must satisfy \(d(\varphi)\leq a(\kappa)\), making
its truth independent of pre-execution history.
State annotations use the restricted formula type
\(\Phi_{\mathsf{st}}\triangleq
\{\varphi\in\Phi\mid\mathsf{NoCurIn}(\varphi)\}\), with
\(U:Q\to\Phi_{\mathsf{st}}\), where \(\mathsf{NoCurIn}\) excludes every current
read \(v@0\) with
\(v\in\mathsf{In}\); current reads of other variables and past reads of inputs
remain admissible. We write \(U_q\) for the annotation of \(q\). Source
programs cannot annotate \(q_0\), so normalization fixes \(U_{q_0}=\top\). Write
\(\Phi_{\mathsf{in}}\triangleq\{\varphi\in\Phi\mid
\forall(v@k)\in\mathsf{reads}(\varphi).\,v\in\mathsf{In}\}\).
For obligation construction, the automata are supplied as finite edge lists
\(E_R\), labelled by \(\Phi_{\mathsf{in}}\), and \(E_G\), labelled by
\(\Phi\), with relations defined by
\[
\begin{aligned}
\delta_R(o,r,r')&\Longleftrightarrow
 \exists\alpha.\ (r,\alpha,r')\in E_R\land
 H_{\mathsf e}(o)\models\alpha,\\
\delta_G(o,g,g')&\Longleftrightarrow
 \exists\gamma.\ (g,\gamma,g')\in E_G\land
 H_{\mathsf p}(o)\models\gamma.
\end{aligned}
\]
Thus automaton guards and generated assertions share the same formula
semantics, while assumptions and guarantees read the entry and post endpoint,
respectively. Input stability and tick completion give
\(H_{\mathsf e}(o)\models\alpha\Longleftrightarrow
H_{\mathsf p}(o)\models\alpha\) for \(\alpha\in\Phi_{\mathsf{in}}\), justifying
the unchanged assumption guard in propagated post-state records.
Three non-uniform maps transport assertions between tick entry, post-state, and
next entry:
\[
\begin{array}{lll}
\mathsf E(v@0)=v@0\ (v\in\mathsf{In}),
&\mathsf E(v@0)=v@1\ (v\notin\mathsf{In}),
&\mathsf E(v@(k+1))=v@(k+1),\\
\mathsf F(v@0)=v@1\ (v\in\mathsf{In}),
&\mathsf F(v@0)=v@0\ (v\notin\mathsf{In}),
&\mathsf F(v@(k+1))=v@(k+2),\\
\mathsf B(v@0)=v@0,
&\mathsf B(v@(k+1))=v@k.&
\end{array}
\tag{2}\label{eq:endpoint-shifts}
\]
Their extension to formulas is structural. \(\mathsf E\) moves an
entry fact to the post-state frame, \(\mathsf F\) moves a
proved post-state fact to the next entry, and \(\mathsf B\) moves a destination
annotation into a postcondition. For a tick with post-store
\(\rho'=c(\rho_{m,\iota})\), let \(H_{\mathsf e}\) and \(H_{\mathsf p}\) denote the entry and
post-state histories. They agree with \(H_m\) at positive depths, with
\(H_{\mathsf e}(v,0)=\rho_{m,\iota}(v)\) and \(H_{\mathsf p}(v,0)=\rho'(v)\). Under the side
conditions stated below,
the endpoint transports satisfy
\[
H_{\mathsf p}\models\mathsf E\varphi\Longleftrightarrow H_{\mathsf e}\models\varphi;\quad
H_{m'}\models\mathsf F\varphi\Longleftrightarrow
H_{\mathsf p}\models\varphi;\quad
H_{\mathsf p}\models\mathsf B\varphi\Longleftrightarrow H_{m'}\models\varphi.
\]
The first equivalence requires coherence of \(m\), input stability, and
completion from \(m\) to \(m'\). The second requires only completion; the
third additionally requires \(\varphi\) to contain no current-input read and
therefore applies to every \(U_q\) by its type. These maps preserve
availability: a post-state record available at age \(a\) yields an
\(\mathsf F\)-image available at age \(a+1\) along that route.

Finite summaries identify the product cases handled by local triples. For
\(x=((q,m),r,g)\), write \(q(x)=q\), \(m(x)=m\), and
\(\kappa(x)=(q,r,g)\). The generator explores the product-key graph from
\(\kappa_0\), combining outgoing program and automaton edges; write
\(\mathsf{KReach}(\kappa)\) for membership in this explored fragment. It groups
cases by a key \(\kappa_\sigma=(q_\sigma,r_\sigma,g_\sigma)\), a core transition
\(\tau_\sigma=(q_\sigma,b_{\tau_\sigma},c_{\tau_\sigma},q'_\sigma)\), and an assumption edge
\(e^R_\sigma=(r_\sigma,\alpha_\sigma,r'_\sigma)\). The finite list \(\Sigma\)
contains exactly the triples for which the source key is structurally reachable
with non-bad assumption and guarantee components, the assumption destination is
non-bad, and the guarantee source has an outgoing edge.

To interpret the Hoare triple generated for a summary, we separate the selected
command body from the guards placed in its precondition. For
\(x=((q,m),r,g)\), \(s'=(q',m')\), and
\(o=(\iota,u,H_{\mathsf e},H_{\mathsf p})\), define
\[
\begin{aligned}
\mathsf{Frame}_\sigma(x,o,s')\Longleftrightarrow{}&
 \kappa(x)=\kappa_\sigma\land\tau_\sigma\in T\land e^R_\sigma\in E_R
 \land q'=q'_\sigma\land\neg\mathsf{bad}_R(r'_\sigma)\\
&{}\land\mathsf{InputStable}(\rho_{m,\iota},c_{\tau_\sigma}(\rho_{m,\iota}))
 \land\mathsf{Complete}(m,c_{\tau_\sigma}(\rho_{m,\iota}),m')\\
&{}\land u=\rho_{m'}|_{\mathsf{Out}}
 \land H_{\mathsf e}=H_{\mathsf e}^{m,\iota}
 \land H_{\mathsf p}=H_{\mathsf p}^{m,m'} .
\end{aligned}
\]
\(\mathsf{Frame}_\sigma\) is the guard-independent semantic frame of that
triple: it fixes the summary context, executes \(c_{\tau_\sigma}\), and records
the program destination and tick observation. It deliberately omits the program
and assumption-edge guards. Guard correspondence represents these omitted
guards by historical formulas: each summary supplies
\(B_\sigma\in\Phi\) such that, under
\(\mathsf{Frame}_\sigma(x,o,s')\),
\(H_{\mathsf e}(o)\models B_\sigma\Longleftrightarrow
\rho_{m,\iota}\in b_{\tau_\sigma}\); the assumption edge label
\(\alpha_\sigma\) is interpreted directly on \(H_{\mathsf e}(o)\).
For an edge \(e=(g_\sigma,\gamma_{\sigma,e},g'_e)\in E_G\), let
\(\mathsf{dst}_\sigma(s',e)=(s',r'_\sigma,g'_e)\). The finite safe and bad
case lists classify outgoing edges by destination:
\[
\begin{aligned}
C_\sigma^+&\triangleq
\{e=(g_\sigma,\gamma,g')\in E_G\mid\neg\mathsf{bad}_G(g')\},\\
C_\sigma^-&\triangleq
\{e=(g_\sigma,\gamma,g')\in E_G\mid\mathsf{bad}_G(g')\}.
\end{aligned}
\]
For each such edge, let
\(G_{\sigma,e}(o)\triangleq H_{\mathsf p}(o)\models\gamma_{\sigma,e}\); the distinction between
\(C_\sigma^+\) and \(C_\sigma^-\) determines whether that guard is propagated
or excluded. Write
\(E^\pm_\sigma(o)=\{e\in C^\pm_\sigma\mid G_{\sigma,e}(o)\}\)
for the compatible safe or bad successor cases, and let
\(\mathsf{Target}^+=\mathsf{Active}\) and
\(\mathsf{Target}^-=\mathsf{BadG}\), and write
\(\mathsf{KActive}(x)=\mathsf{KReach}(\kappa(x))\land\mathsf{Active}(x)\).
A summary represents an actual product case only when its program and
assumption guards hold. Define
\[
  \mathsf{Exec}_\sigma(x,o,s')
  \triangleq
  \mathsf{Frame}_\sigma(x,o,s')\land
  H_{\mathsf e}(o)\models B_\sigma\land
  H_{\mathsf e}(o)\models\alpha_\sigma.
\tag{3}\label{eq:summary-exec}
\]
For a product state \(y=(s,r,g)\), write \(\mathsf{prog}(y)=s\). Using
\eqref{eq:summary-exec}, the summaries characterize, for every \(x,o,y\) and
\(\epsilon\in\{+,-\}\), the corresponding product edges in both directions:
\[
\begin{aligned}
\mathsf{KActive}(x)\land x\xRightarrow{o}y
 \land\mathsf{Target}^{\epsilon}(y)
&\Longleftrightarrow
\exists\sigma\in\Sigma,\ e\in E^\epsilon_\sigma(o).\\
&\quad y=\mathsf{dst}_\sigma(\mathsf{prog}(y),e)
 \land\mathsf{Exec}_\sigma(x,o,\mathsf{prog}(y)).
\end{aligned}
\tag{4}\label{eq:summary-exactness}
\]
The forward implication decomposes an observed product step and the reverse
implication reconstructs one. Both are proved for the finite symbolic
summaries, not assumed by the generated-obligation theorem.

The certificate \eqref{eq:invariant-certificate} combines two persistent facts: the
user annotation \(U_q=U(q)\in\Phi_{\mathsf{st}}\) of each program control state,
and a generated product characteristic recording how a selected product state
was entered. For
each safe edge \(e\), let \(\beta_\sigma\in\Phi\) be a post-state summary of
the body and define
\[
 D_{\sigma,e}^{\mathsf{post}}
 \triangleq
 \mathsf E(U_{q_\sigma})\land\mathsf E(B_\sigma)
 \land\alpha_\sigma\land\gamma_{\sigma,e}
              \land\beta_\sigma .
\tag{5}\label{eq:incoming-contribution}
\]
These guarded records form \(\mathsf{Contrib}^{\mathsf{post}}\) in
\Cref{sec:tool-pipeline}; the first conjunct retains source-annotation facts
needed by later historical reads. The implementation obtains \(\beta_\sigma\) by symbolic execution of
straight-line assignments and frame equalities; values assigned by unsupported
compound statements are omitted. This weakening remains sound: the safe
obligation executes the actual body and must prove
\eqref{eq:incoming-contribution} afterward. For \(e\in C_\sigma^+\), write
\(\kappa_{\sigma,e}\triangleq(q'_\sigma,r'_\sigma,g'_e)\) for the combined
destination key. For each product key \(\kappa\), define
\[
 L_\kappa\triangleq
 \{(\sigma,e)\mid \sigma\in\Sigma,\ e\in C_\sigma^+,\ 
 \kappa_{\sigma,e}=\kappa\}.
\]
Let \(\mathsf{Start}_\kappa\) be \(\top\) when \(\kappa=\kappa_0\) and
\(\bot\) otherwise. The generator enumerates \(L_\kappa\) exactly and defines
\[
 \chi_\kappa^{\mathsf{entry}}\triangleq
 \mathsf F\!\left(
   \mathsf{Start}_\kappa\lor
   \bigvee_{(\sigma,e)\in L_\kappa}
     D_{\sigma,e}^{\mathsf{post}}\right).
\tag{6}\label{eq:product-characteristic}
\]
Each \(D_{\sigma,e}^{\mathsf{post}}\) is established by one safe predecessor
obligation, whereas \(\chi_\kappa^{\mathsf{entry}}\) is their shifted
disjunction, assumed by obligations whose source key is \(\kappa\). It is
\(\mathsf{Incoming}(\kappa)\) in \Cref{sec:tool-pipeline}.
The startup disjunct makes \(\chi_{\kappa_0}^{\mathsf{entry}}=\top\).
Accordingly,
\(D_{\sigma,e}^{\circ}\triangleq D_{\sigma,e}^{\mathsf{post}}\lor
\mathsf{Start}_{\kappa_{\sigma,e}}\) is \(D_{\sigma,e}^{\mathsf{post}}\)
for a noninitial destination and \(\top\) for the initial destination, which
is also reachable at startup without a predecessor.
The construction yields pre- and postconditions that preserve the certificate
or exclude guarantee-bad successors. The formulas emitted for a summary are
\[
\begin{aligned}
R_\sigma \triangleq{}\;& U_{q_\sigma}\land\chi^{\mathsf{entry}}_{\kappa_\sigma}
 \land B_\sigma\land\alpha_\sigma,
\qquad
Q_\sigma^- \triangleq
 \bigwedge_{e\in C_\sigma^-}\neg\gamma_{\sigma,e},\\
Q_\sigma^+ \triangleq{}\;&
 \biggl(\bigvee_{e\in C_\sigma^+}\gamma_{\sigma,e}\biggr)
 \land\!\bigwedge_{e\in C_\sigma^+}\!\Bigl(
 \gamma_{\sigma,e}\Rightarrow
 [\mathsf B(U_{q'_\sigma})\land
   D_{\sigma,e}^{\circ}]\Bigr).
\end{aligned}
\tag{7}\label{eq:generated-postconditions}
\]
In \Cref{sec:tool-pipeline}'s notation, the safe disjunction
is \(\mathsf{SomeSafe}\), the guarded conjunction establishes
\(\mathsf{Inv}^{\mathsf{post}}=\mathsf B(U)\) and
\(\mathsf{Contrib}^{\mathsf{post}}\) from \(D^\circ\), and \(Q^-_\sigma\) is
\(\mathsf{NoBad}\). The semantic \(\mathsf{Coherent}_{\mathsf{In}}\) premise
corresponds to the finite \(\mathsf{EntryCoherence}\) equalities emitted by the
prototype.
\paragraph{Generated-family conditions.}
Frontier certificates make generated formulas independent of unavailable
history, and each \(B_\sigma\) denotes its executable guard. For nonempty
\(C_\sigma^\epsilon\), the family contains
\[
 \{R_\sigma\}\ c_{\tau_\sigma}\ \{Q_\sigma^\epsilon\},
 \qquad \epsilon\in\{+,-\},\quad C_\sigma^\epsilon\ne\varnothing .
\tag{8}\label{eq:generated-obligations}
\]
A triple is valid when, for every \(x,o,s'\),
\(\mathsf{Frame}_\sigma(x,o,s')\),
\(\mathsf{Coherent}_{\mathsf{In}}(m(x))\), and
\(H_{\mathsf e}(o)\models R_\sigma\) imply
\(H_{\mathsf p}(o)\models Q_\sigma^\epsilon\). Validity uses
\(\mathsf{Frame}_\sigma\), rather than \(\mathsf{Exec}_\sigma\), because
\(B_\sigma\) and \(\alpha_\sigma\) occur explicitly in \(R_\sigma\), exactly as
in the emitted Why3 precondition. Why3 total-correctness justifies the total
command denotation used by \(\mathsf{Frame}_\sigma\).
These conditions and valid safe and bad triples constitute generated-obligation
validity. Endpoint transport and summary exactness are proved by construction;
source-formula characterization and normalized-core assumptions are used only
for trace soundness and reactivity.

\begin{theorem}[Generated-obligation soundness]
\label{thm:generated-obligation-soundness}
Generated-obligation validity for \eqref{eq:generated-obligations} entails
\(\mathsf{ContractValid}(\mathcal P,A_R,A_G)\).
\end{theorem}

Through \eqref{eq:generated-postconditions}, safe obligations propagate proved
records to build the invariant; bad ones use it to exclude violating successors.
The certificate is:
\[
 I(x)\triangleq
 \mathsf{KReach}(\kappa(x))
 \land\bigl(H_{m(x)}\models U_{q(x)}\bigr)
 \land\mathsf{Coherent}_{\mathsf{In}}(m(x))
 \land\bigl(H_{m(x)}\models\chi^{\mathsf{entry}}_{\kappa(x)}\bigr).
\tag{9}\label{eq:constructed-certificate}
\]
The \(\mathsf{KReach}\) conjunct of \eqref{eq:constructed-certificate} records finite product exploration in the
metaproof and is not emitted to the backend. By
\eqref{eq:summary-exactness}, every active safe product edge from such a key has
a generated case. Its obligation establishes the destination encoding of \(U\)
and the corresponding incoming disjunct at every noninitial destination, while
\(\mathsf{Complete}\) re-establishes coherence; the endpoint shifts
\eqref{eq:endpoint-shifts}
therefore establish \(I\) at the successor, giving \textsc{Pres}. At a
source with a bad successor, its characteristic
\eqref{eq:product-characteristic} supplies the previously proved incoming
disjunction and the bad obligation gives \textsc{Excl}. The initial
\(U_{q_0}=\top\), initial-memory coherence, and the non-bad guarantee initial
state give \textsc{Init};
\Cref{thm:invariant-soundness} then applies.

Contract validity is a product-level bad-prefix property. Completeness of
\(A_G\) also ensures a non-bad guarantee successor whenever a reachable program
state advances and the assumption remains non-bad.

\paragraph{Trace-characterization assumptions.}
Let \(R\) and \(G\) denote source trace properties and \(\pi_{\mathsf{in}}\)
select inputs pointwise; for every observation word \(w\), require
\[
 \pi_{\mathsf{in}}(w)\models R\Longleftrightarrow w\Vdash A_R,
 \qquad
 w\models G\Longleftrightarrow w\Vdash A_G.
\]
For an existing program run satisfying \(R\), the first equivalence supplies
assumption coverage and excludes assumption-bad prefixes. Product soundness
excludes guarantee-bad prefixes, and completeness of \(A_G\) supplies the
coverage needed by the second equivalence. Hence
\Cref{thm:generated-obligation-soundness} has the trace-level corollary
\[
  \forall\varrho\in\mathsf{Runs}(\mathcal P).\quad
  \pi_{\mathsf{in}}(\mathsf{tr}(\varrho))\models R
  \Rightarrow\mathsf{tr}(\varrho)\models G,
\tag{10}\label{eq:trace-soundness}
\]

Trace soundness remains conditional on a run; the normalized-core assumptions
ensure its existence and uniqueness for each input stream.
\begin{theorem}[Reactive contract correctness]
\label{thm:reactive-contract-correctness}
Validity of the generated obligations entails that every
\(\bar\iota\models R\) induces a unique
\(\varrho\in\mathsf{Runs}(\mathcal P,\bar\iota)\) satisfying
\(\mathsf{tr}(\varrho)\models G\).
\end{theorem}
Coverage, input preservation, and total statement denotations construct a run;
determinism makes it unique, and \eqref{eq:trace-soundness} establishes \(G\).

\subsection{Formalization boundary}

The formalized core connects normalized ticks and histories to the edge-labelled
product, proves the exactness of finite summaries, incoming characteristics, and
historical shifts, and derives reactive correctness from generated Hoare triples;
source syntax and construction enter only through their denotations.
Trust assumptions cover source-normalization correspondence (priority, fallback
coverage, input discipline, and expression and statement denotations), automata
translation, OCaml enumeration, Why3 encoding and solver answers, lowering
\texttt{pre}/\texttt{pre\_k} reads to backend parameters, history-depth checks,
and the link from backend total correctness to total denotations.
At the automata boundary, \spot checks safety and returns finite complete
automata whose rejected prefixes characterize the source formulas \cite{spot}.
The adapter merges rejecting states into a sink; the product builder rejects
bad initial states and checks \(A_R\)-absorption. Thus finiteness,
characterization, non-bad initials, \(A_G\)-completeness, and
\(A_R\)-absorption are supplied or checked externally.

\section{Related Work}
\label{sec:related-work}

\kairos combines synchronous programming, automata-theoretic verification,
assertional proof, and SMT-backed deduction. We compare related workflows by
their temporal proof boundary: a synchronous observer, symbolic transition
system, abstract product, or generated deductive interface.

\emph{Synchronous contracts and symbolic engines.}
\kairos follows the logical-instant model of Lustre and Esterel
\cite{berry1992esterel,caspi1987lustre,benveniste2003synchronous} and
SCADE-style explicit modes \cite{scade6}. Observer-based Lustre workflows encode
temporal requirements into synchronous state and prove the resulting Boolean
streams invariant \cite{halbwachs-lagnier-raymond-observers}; Kind~2 and JKind
symbolically verify Lustre programs and contracts \cite{kind2,jkind}, while AGREE
compositionally analyzes architectural assume--guarantee contracts using engines
such as JKind \cite{agree,jkind}. At this verification boundary, temporal intent
must already be encoded as observer state or stream/contract equations, manually
or by an upstream frontend. \kairos instead accepts temporal assumptions and
guarantees at source level and generates automata carrying only their finite
temporal control. Imperative bodies and data are handled separately by the
deductive backend, so support for data theories and command constructs depends
on that backend rather than on a symbolic temporal engine.

\emph{Automata and assertional temporal proof.}
The automata-theoretic treatment of temporal specifications is classical
\cite{vardi-wolper-1986,vardi-wolper-1994}. The safety/liveness distinction of
Alpern and Schneider explains why the present construction can reduce safety to
exclusion of bad product states \cite{alpern-schneider}. Their assertional
method already derives local obligations over a program--automaton structure,
using invariants for safety and variants for progress
\cite{alpern-schneider-1989}. We therefore do not claim either the
automata-product principle or its assertional reading as new. The contribution
is the synchronous source-to-obligation boundary needed to realize the safety
case in an executable tool. Edge observations must align automata with sampled
program ticks; assumptions and guarantees impose different obligations;
historical assertions must be transported between entry and exit endpoints;
current inputs cannot become persistent state facts; safe predecessor
information must be propagated; and finite summaries must retain every
compatible successor.

\emph{Deductive verification of reactive programs.}
Manna--Pnueli and STeP combine temporal specifications with assertions,
verification diagrams, abstraction, and deduction
\cite{manna-pnueli-safety,step-update-1999}, leaving temporal reasoning within
the proof problem. Lustre/PVS methods are closer: they use continuous induction
to reduce safety properties already encoded as Lustre observers to scalar PVS
obligations \cite{lustre-pvs-methodology,lustre-pvs-obligation-generator}. Gesell and
Schneider give Hoare rules for synchronous macro-steps
\cite{gesell-schneider-synchronous-hoare}, while recent Why3 work translates
Ladder code and state-transition diagrams into WhyML tasks
\cite{cousineau-ladder-std-2025}. Compared with these approaches, \kairos
translates temporal assume--guarantee contracts through safety automata into
ordinary first-order pre/postconditions around source commands. Product
summaries, incoming characteristics, and historical shifts account for temporal
meaning before Why3 and SMT solvers discharge the resulting data and command
obligations \cite{why3,z3}.

\emph{Mechanization, compilation, and monitoring.}
The normalized imperative core resembles intermediate forms used in synchronous
compilation, notably object-based code (OBC) in Bourke et al.'s verified Lustre compiler
\cite{bourke-etal-pldi2017-lustre}. Whereas that work proves compilation
correctness, we mechanize the correctness of temporal-contract obligation
construction over such a core \cite{rocq-prover}. Runtime
verification, including Copilot, compiles temporal intent into state maintained
during execution \cite{copilot}. \kairos instead uses automata statically to
organize obligations over symbolic source ticks, rather than monitoring a single
execution.

\section{Conclusion}
\label{sec:conclusion}

We presented \kairos, a source-level toolchain translating
assume--guarantee safety contracts into automata synchronized with normalized
program ticks and then into first-order preservation and guarantee-bad exclusion
obligations around command bodies. Finite summaries retain every compatible
guarantee successor; incoming characteristics and explicit historical shifts
propagate proved safe cases into later preconditions. This exact synchronous
interface realizes the classical
reduction of temporal safety to product invariance. The \rocq
mechanization establishes that valid generated initial and transition obligations
imply contract validity and, with coverage, input preservation, determinism, and
total command denotations, that every assumption-satisfying input trace induces
a unique guarantee-satisfying execution. Future directions include proving
non-spuriousness, which would replace
regression-based confidence with a proof that supplied annotations yield suitable
local assertions for valid contracts, and reducing trust through checkable
certificates linking generated summaries, backend obligations, Why3 total
correctness, and command denotations. Further directions are AGREE-style
compositional assume--guarantee verification of synchronous-node networks
\cite{agree}, separation of private generated temporal state from reusable
interfaces, and ranking or variant obligations beyond safety
\cite{alpern-schneider-1989}.

\bibliographystyle{splncs04}
\bibliography{refs}

@article{alpern-schneider,
  author  = {Alpern, Bowen and Schneider, Fred B.},
  title   = {Defining Liveness},
  journal = {Information Processing Letters},
  year    = {1985},
  volume  = {21},
  number  = {4},
  pages   = {181--185},
  doi     = {10.1016/0020-0190(85)90056-0}
}

@article{alpern-schneider-1989,
  author  = {Alpern, Bowen and Schneider, Fred B.},
  title   = {Verifying Temporal Properties without Temporal Logic},
  journal = {ACM Transactions on Programming Languages and Systems},
  year    = {1989},
  volume  = {11},
  number  = {1},
  pages   = {147--167},
  doi     = {10.1145/59287.62028}
}

@techreport{agree,
  author      = {Backes, John and Cofer, Darren and Amundson, Isaac},
  title       = {The Assume Guarantee Reasoning Environment with Application to an Unmanned Helicopter},
  institution = {Loonwerks},
  year        = {2021},
  url         = {https://loonwerks.com/publications/pdf/backes2021techreport.pdf}
}

@incollection{step-update-1999,
  author    = {Berghammer, Rudolf and Lakhnech, Yassine and Manna, Zohar and Bj{\o}rner, Nikolaj S. and Browne, Anca and Col{\'o}n, Michael and Finkbeiner, Bernd and Pichora, Mark and Sipma, Henny B. and Uribe, Tom{\'a}s E.},
  title     = {An Update on {STeP}: Deductive-Algorithmic Verification of Reactive Systems},
  booktitle = {Tool Support for System Specification, Development and Verification},
  year      = {1999},
  pages     = {174--188},
  publisher = {Springer Vienna},
  address   = {Vienna},
  doi       = {10.1007/978-3-7091-6355-9_13}
}

@article{benveniste2003synchronous,
  author  = {Benveniste, Albert and Caspi, Paul and Edwards, Stephen A. and Halbwachs, Nicolas and Le Guernic, Paul and de Simone, Robert},
  title   = {The Synchronous Languages 12 Years Later},
  journal = {Proceedings of the IEEE},
  volume  = {91},
  number  = {1},
  pages   = {64--83},
  year    = {2003},
  doi     = {10.1109/JPROC.2002.805826}
}

@article{berry1992esterel,
  author  = {Berry, G{\'e}rard and Gonthier, Georges},
  title   = {The Esterel Synchronous Programming Language: Design, Semantics, Implementation},
  journal = {Science of Computer Programming},
  volume  = {19},
  number  = {2},
  pages   = {87--152},
  year    = {1992},
  doi     = {10.1016/0167-6423(92)90005-V}
}

@inproceedings{bourke-etal-pldi2017-lustre,
  author    = {Bourke, Timothy and Brun, L{\'e}lio and Dagand, Pierre-\'{E}variste and Leroy, Xavier and Pouzet, Marc and Rieg, Lionel},
  title     = {A Formally Verified Compiler for Lustre},
  booktitle = {Proceedings of the 38th ACM SIGPLAN Conference on Programming Language Design and Implementation},
  year      = {2017},
  pages     = {586--601},
  publisher = {ACM},
  address   = {New York, NY, USA},
  doi       = {10.1145/3062341.3062358}
}

@inproceedings{caspi1987lustre,
  author    = {Caspi, Paul and Pilaud, Daniel and Halbwachs, Nicolas and Plaice, John A.},
  title     = {{LUSTRE}: A Declarative Language for Real-Time Programming},
  booktitle = {Proceedings of the 14th ACM SIGACT-SIGPLAN Symposium on Principles of Programming Languages},
  pages     = {178--188},
  year      = {1987},
  publisher = {ACM},
  address   = {New York, NY, USA},
  doi       = {10.1145/41625.41641}
}

@article{copilot,
  author  = {Pike, Lee and Wegmann, Nis and Niller, Sebastian and Goodloe, Alwyn},
  title   = {Copilot: Monitoring Embedded Systems},
  journal = {Innovations in Systems and Software Engineering},
  year    = {2013},
  volume  = {9},
  number  = {4},
  pages   = {235--255},
  doi     = {10.1007/s11334-013-0223-x}
}

@inproceedings{gesell-schneider-synchronous-hoare,
  author    = {Gesell, Manuel and Schneider, Klaus},
  title     = {A Hoare Calculus for the Verification of Synchronous Languages},
  booktitle = {Proceedings of the 6th Workshop on Programming Languages Meets Program Verification},
  year      = {2012},
  pages     = {37--48},
  publisher = {ACM},
  address   = {New York, NY, USA},
  doi       = {10.1145/2103776.2103782}
}

@inproceedings{kind2,
  author    = {Champion, Adrien and Mebsout, Alain and Sticksel, Christoph and Tinelli, Cesare},
  title     = {The {Kind 2} Model Checker},
  booktitle = {Computer Aided Verification},
  series    = {Lecture Notes in Computer Science},
  volume    = {9780},
  year      = {2016},
  pages     = {510--517},
  publisher = {Springer},
  address   = {Cham},
  doi       = {10.1007/978-3-319-41540-6_29}
}

@inproceedings{lustre-pvs-methodology,
  author    = {Bensalem, Saddek and Caspi, Paul and Dumas, C{\'e}cile and Parent-Vigouroux, Catherine},
  title     = {A Methodology for Proving Control Systems with Lustre and {PVS}},
  booktitle = {Dependable Computing for Critical Applications---7},
  series    = {Dependable Computing and Fault Tolerant Systems},
  volume    = {12},
  year      = {1999},
  pages     = {89--107},
  publisher = {IEEE Computer Society},
  address   = {San Jose, CA, USA},
  doi       = {10.1109/DCFTS.1999.814291}
}

@inproceedings{lustre-pvs-obligation-generator,
  author    = {Canovas-Dumas, C{\'e}cile and Caspi, Paul},
  title     = {A {PVS} Proof Obligation Generator for Lustre Programs},
  booktitle = {Logic for Programming and Automated Reasoning},
  series    = {Lecture Notes in Artificial Intelligence},
  volume    = {1955},
  year      = {2000},
  pages     = {179--188},
  publisher = {Springer},
  address   = {Berlin, Heidelberg},
  doi       = {10.1007/3-540-44404-1_12}
}

@book{manna-pnueli-safety,
  author    = {Manna, Zohar and Pnueli, Amir},
  title     = {Temporal Verification of Reactive Systems: Safety},
  publisher = {Springer},
  address   = {New York, NY, USA},
  year      = {1995},
  doi       = {10.1007/978-1-4612-4222-2}
}

@misc{rocq-prover,
  author       = {{The Rocq Development Team}},
  title        = {The Rocq Prover},
  year         = {2025},
  howpublished = {Software, Version 9.1.0},
  url          = {https://rocq-prover.org/}
}

@inproceedings{halbwachs-lagnier-raymond-observers,
  author    = {Halbwachs, Nicolas and Lagnier, Fabienne and Raymond, Pascal},
  title     = {Synchronous Observers and the Verification of Reactive Systems},
  booktitle = {Algebraic Methodology and Software Technology},
  series    = {Workshops in Computing},
  year      = {1993},
  pages     = {83--96},
  publisher = {Springer},
  address   = {London, UK},
  doi       = {10.1007/978-1-4471-3227-1_8}
}

@inproceedings{scade6,
  author    = {Cola{\c{c}}o, Jean-Louis and Pagano, Bruno and Pouzet, Marc},
  title     = {{SCADE} 6: A Formal Language for Embedded Critical Software Development},
  booktitle = {Theoretical Aspects of Software Engineering},
  year      = {2017},
  pages     = {1--11},
  publisher = {IEEE Computer Society},
  address   = {Los Alamitos, CA, USA},
  doi       = {10.1109/TASE.2017.8285623}
}

@inproceedings{spot,
  author    = {Duret-Lutz, Alexandre and Renault, Etienne and Colange, Maximilien and Renkin, Florian and Aisse, Alexandre Gbaguidi and Schlehuber-Caissier, Philipp and Medioni, Thomas and Martin, Antoine and Dubois, J{\'e}r{\^o}me and Gillard, Cl{\'e}ment and Lauko, Henrich},
  title     = {From Spot 2.0 to Spot 2.10: What's New?},
  booktitle = {Computer Aided Verification},
  series    = {Lecture Notes in Computer Science},
  volume    = {13372},
  year      = {2022},
  pages     = {174--187},
  publisher = {Springer},
  address   = {Cham},
  doi       = {10.1007/978-3-031-13188-2_9}
}

@inproceedings{vardi-wolper-1986,
  author    = {Vardi, Moshe Y. and Wolper, Pierre},
  title     = {An Automata-Theoretic Approach to Automatic Program Verification},
  booktitle = {Proceedings of the First Symposium on Logic in Computer Science},
  year      = {1986},
  pages     = {332--344},
  publisher = {IEEE Computer Society},
  address   = {Los Alamitos, CA, USA}
}

@article{vardi-wolper-1994,
  author  = {Vardi, Moshe Y. and Wolper, Pierre},
  title   = {Reasoning about Infinite Computations},
  journal = {Information and Computation},
  year    = {1994},
  volume  = {115},
  number  = {1},
  pages   = {1--37},
  doi     = {10.1006/inco.1994.1092}
}

@inproceedings{why3,
  author    = {Filli{\^a}tre, Jean-Christophe and Paskevich, Andrei},
  title     = {Why3 --- Where Programs Meet Provers},
  booktitle = {Programming Languages and Systems},
  series    = {Lecture Notes in Computer Science},
  volume    = {7792},
  year      = {2013},
  pages     = {125--128},
  publisher = {Springer},
  address   = {Berlin, Heidelberg},
  doi       = {10.1007/978-3-642-37036-6_8}
}

@inproceedings{z3,
  author    = {de Moura, Leonardo and Bj{\o}rner, Nikolaj},
  title     = {{Z3}: An Efficient {SMT} Solver},
  booktitle = {Tools and Algorithms for the Construction and Analysis of Systems},
  series    = {Lecture Notes in Computer Science},
  volume    = {4963},
  year      = {2008},
  pages     = {337--340},
  publisher = {Springer},
  address   = {Berlin, Heidelberg},
  doi       = {10.1007/978-3-540-78800-3_24}
}

@inproceedings{jkind,
  author    = {Gacek, Andrew and Backes, John and Whalen, Mike and Wagner, Lucas and Ghassabani, Elaheh},
  title     = {The {JKind} Model Checker},
  booktitle = {Computer Aided Verification},
  series    = {Lecture Notes in Computer Science},
  volume    = {10982},
  year      = {2018},
  pages     = {20--27},
  publisher = {Springer},
  address   = {Cham},
  doi       = {10.1007/978-3-319-96142-2_3}
}

@techreport{cousineau-ladder-std-2025,
  author      = {Cousineau, Denis and Inoue, Hiroaki and March{\'e}, Claude and Faissole, Florian and Mentr{\'e}, David},
  title       = {Conformity of Ladder Programs against State Transition Diagrams},
  institution = {Inria},
  number      = {RR-9584},
  year        = {2025},
  month       = mar,
  url         = {https://inria.hal.science/hal-05001069}
}

\end{document}